\documentclass{article}
\usepackage[utf8]{inputenc}
\usepackage{mathtools}
\usepackage{subfigure}
\usepackage{authblk}

\newcommand{\tobs}{\ensuremath{t_{\mathrm{obs}}}}
\usepackage{geometry}
\title{Per-event significance indicator to visualise significant events}

\author[1]{Nicholas Wardle}
\affil[1]{Imperial College London, n.wardle09@ic.ac.uk}
\date{November 2018}

\begin{document}

\maketitle

\abstract{In this note, an alternative for presenting the distribution of `significant' events in searches 
for new phenomena is described. The alternative is based on probability density functions used in the 
evaluation of the `significance' of an observation, rather than the typical ratio of signal to background. 
The method is also applicable to searches that use unbinned data, for which the concept of signal to background can be ambiguous. In the case of simple searches using binned data, this method reproduces the familiar quantity $\log(s/b)$, when the signal to background ratio is small.}

\section{Introduction}

Discoveries of new phenomena in particle physics typically involve the observation of an excess or deficit of events with respect to that 
which the current, best theories predict. The discovery of the Z boson at the UA1 and UA2 experiments~\cite{UA1,UA2}, 
was announced on the observation of only a handful of events, which formed clear excesses in mass distribution of the final state particles. 
In modern high energy particle physics colliders  the 
experimental signature for new phenomena is less clean, and requires the use of 
sophisticated statistical techniques to detect.
In the Higgs boson search at the Tevatron, a combination of different decay channels using data from CDF and D0 was used to extract the Higgs signal. This combination pointed to an excess of events with an overall significance of 3 standard deviations~\cite{TEV_H}. The data were split into multiple bins, depending on the kinematics of 
the events or additional particles present in the events, with differing sensitivities to the Higgs boson signal.
Due to the fact that each of these categories contributes to the overall significance of the result, it was not  
possible to `see' a clear signal in a single bin or distribution of events. The events were therefore binned in a quantity related to the signal to background ratio so that the signal can be seen in the data in a single figure. The distribution of the quantity $\text{log}_{10}(s/b)$ from the search bins was  used to demonstrate the varying sensitivity of the different bins in the analysis and to indicate bins which most contributed to the significance of the excess (see figure 1. from reference~\cite{TEV_H}). 

After the discovery of the Higgs boson at ATLAS and CMS~\cite{ATLAS_H,CMS_H_1,CMS_H_2}, the search for rarer modes of production became the focus. Most recently, the analysis which lead to the announcement of the discovery of the ttbar Higgs production (ttH) mode~\cite{CMSttH}, also utilized 
many different event categories, from a combination of several decay channels. Again, the same quantity has been used to visualise the excess of the newly observed process (see figure 3. of reference~\cite{CMSttH}).

There are however three issues with these methods:
\begin{enumerate} 
\item{The signal to background ratio (or the log of it) is \textit{not} exactly the figure of merit which 
contributes to the overall significance of the signal. Even when the two hypotheses represent the `background only' and `signal plus background', the background component may not always be the same in both -- for example, in the presence of nuisance parameters whose values are obtained from a fit to data. This is also true in cases where interference between the signal and background can result in a reduction of events in the presence of a signal.}
\item{For hypotheses tests which aim to compare two different scenarios for an observed signal, the notion of signal to background ratios is not an appropriate 
figure of merit to judge the compatibility of the signal with either hypothesis.  }
\item{The concept of a signal to background ratio is inherently a `binned' concept. In analyses that use \emph{unbinned  data}, the ratio of signal to background is not uniquely defined.}
\end{enumerate}
With these issues in mind, this note outlines a proposal for a new visualization for data, which overcomes these issues. 
Moreover, the method is well suited to current searches for new phenomena being conducted at the LHC, or any other 
experiment which uses a profiled likelihood based approach. This includes studies which compare two hypotheses for an observed signal. 
The method makes use of pseudo-data in order to visualise the compatibility of the data with the two hypotheses  
-- the significance of the result. These pseudo-data can however be substituted by an Asimov dataset~\cite{CCGV} in cases where the search is 
sufficiently complex enough, or the excess sufficiently large enough, to make the use of pseudo-datasets computationally 
prohibitive.  

\section{Likelihoods and significance}

It is useful to review the common method by which the significance of an excess in data is quantified, in the context of a 
particular search for some new phenomena, particularly at the LHC. Typically, the significance, $p$, is calculated as the probability to observe some outcome in the data, assuming some hypothesis $H$, which is at least as discrepant with $H$ as the actual outcome observed,

\begin{equation}
    p = \int_{C_{\tobs}} f\left(t|H\right)dt.
\end{equation}
Here, $t$ is a real valued number known as the `test-statistic`, with distribution $f\left(t|H\right)$ under the hypothesis $H$. The region of integration region $C_{\tobs}$ is usually determined before making the observation $\tobs$. A simple example would be the observation of some decay process, in which $n_{\mathrm{exp}}$ decays are expected under the hypothesis $H$, in a fixed time interval. The test-statistic $t$ in this case is the number of decays and $\tobs=n_{\mathrm{obs}}$, the observed number of decays. If $n_{\mathrm{exp}}$ is small, $f\left(t|H\right)$ is a Poisson distribution with mean parameter $n_{\mathrm{exp}}$ and the region $C_{\tobs}$ is simply $\left\{n:n \geq n_{\mathrm{obs}}\right\}$. Typically, when $p$ is smaller than some threshold value $\alpha$, the hypothesis $H$ is said 
to be rejected at the $100\times(1-\alpha)$\% confidence level (CL)\footnote{Since if $t$ is continuous, the distribution of $p$ is 
uniform under the hypothesis $H$, it is also the case that if $H$ is excluded when $p<\alpha$, then $H$ will be rejected  in a fraction of $\alpha$ of the outcomes, even if $H$ is true. This is known as a type--1 error}.  In particle physics experiments, the test-statistic must be able to summarize the entire data with a single number, for example by defining a likelihood under 
any potential hypothesis $H$. A common choice of test-statistic in particle physics makes use of  
log-likelihood ratios; $t=-2\ln \left(\frac{\mathcal{L}_{a}}{\mathcal{L}_{b}}\right)$, which compare the likelihood under two hypotheses $H_{a}$ and $H_{b}$\footnote{It is common to drop the explicit reference to the data. Throughout this paper therefore, the abbreviation $\mathcal{L}(\text{data}|H_{X}):=\mathcal{L}_{X}$ has been made}. 

Often, systematic uncertainties will be incorporated through the introduction of `nuisance parameters' $\nu$. This means that often a particular hypothesis will be fully specified under a particular set of values for these nuisance parameters, i.e $H\rightarrow H|_{\nu}$. In general this involves either `profiling' the nuisance parameters (performing a fit to the data) to remove the dependence of the hypothesis (and therefore the likelihood) on them.   
To account for these nuisance parameters The likelihood is typically augmented by the inclusion 
of constraint terms (or prior probability densities) $\pi(\nu)$ -- a trivial version of which is a flat prior which makes the nuisance parameter `unconstrained`. The inclusion 
of nuisance parameters presents no complication for defining the per-event significance indicator, adding only a fixed term in each event making this method well suited to incorporating systematic uncertainties. The method is also applicable to any search using a log-likelihood ratio, which does not incorporate systematic uncertainties via nuisance parameters. 
However in order to account for systematic uncertainties, additional toy distributions 
should be generated from alternative scenarios for the systematic variations and the resulting per-event significance indicator distribution  
can be augmented with an `uncertainty band' determined from these toys. The details of this procedure are left to the 
reader as they are beyond the scope of this note.

A very generic likelihood function, often utilized in particle physics, can be expressed as,

\begin{equation}
    \mathcal{L} = \frac{\lambda^{n} e^{-\lambda} }{n!} \cdot \prod_{i=1}^{n} \rho\left(x_{i}|H\right),
\label{eqn:lh}
\end{equation}
where $\rho$ is the probability density function, under the hypothesis $H$ for the observable $x$, and $\int \rho\left(x|H\right)=1$. Here, $x$ can represent one or more discrete or continuous quantities -- e.g. invariant mass, lifetime, number of charged particles, etc. --  which exhibit some separation power between two hypotheses -- commonly these hypothesis represent the presence (or not) of some signal process. Let the `signal plus background' and `background only' hypotheses be $H_{\mathcal{S}}$ and $H_{\mathcal{B}}$, respectively. Then the test-statistic 
becomes, 

\begin{equation}
   t =  -2\ln\left(\frac{\mathcal{L_{B}}}{\mathcal{L_{S}}}\right) = 
   2 \sum_{i=1}^{n}\left[\ln\rho(x_{i}|H_{\mathcal{S}}) - \ln\rho(x|H_{\mathcal{B}}) \right] 
   + 2\left(\lambda_{\mathcal{B}}-\lambda_{\mathcal{S}}\right) 
   + 2n\ln\left(\frac{\lambda_{\mathcal{S}}}{\lambda_{\mathcal{B}}}\right),
\end{equation}
where $\lambda_{\mathcal{S}}$ and $\lambda_{\mathcal{B}}$ correspond to the `signal plus background' and `background only' hypotheses, respectively\footnote{As previously mentioned, this also holds for other hypotheses comparisons, $H_{a}$ vs $H_{b}$, not necessarily between `signal plus background' and `background only'.}.

Typically, values of $t$ which are far away from those expected under $H_{\mathcal{B}}$ will correspond to observations that are \emph{significant}. This is evident from the fact that $p$ will be small when $t_\text{obs}$ lies in the extreme values of the distribution of $t$. A larger than expected\footnote{where expected is defined usually with respect to the background only hypothesis} value of $t_{\mathrm{obs}}$ indicates that the likelihood is much larger under one hypothesis than the other. In certain cases, the value of $t$ will directly correspond to the significance -- e.g. the test-statistic for discovery at the LHC has the asymptotic property that the significance $Z=\sqrt{t}$~\cite{CCGV}, though this is not true in general. 

\section{Per-event significance indicator}

While the test-statistic $t$ is indeed a good indicator of the significance of an observation -- with `unlikely' values of $t$ being used to reject the null hypothesis -- it is not so clear which events in particular are the source of the significant result. 
Assuming equation~\ref{eqn:lh},  for a given event $i$, we can define the quantity, 

\begin{equation}
    \xi = \ln\rho(x_{i}|H_{\mathcal{S}}) - \ln\rho(x_i|H_{\mathcal{B}}) + \frac{1}{n}\left(\lambda_{\mathcal{B}}-\lambda_{\mathcal{S}}\right) + \ln\left(\frac{\lambda_{\mathcal{S}}}{\lambda_{\mathcal{B}}}\right),  
\label{eqn:per-event-si}
\end{equation}
such that $t = 2\sum_{i=1}^{n}\xi_{i}$. This quantity is the \emph{per-event significance indicator}. Events with large, positive values of this quantity contribute most to the overall significance, while events will small values do not contribute. Events with large negative values actually detract from the overall significance and so this quantity shows how some events can reduce the significance of an excess. 

In the case of particle physics searches at the LHC, the likelihood will typically include `nuisance parameters', $\nu$, which are \emph{constrained} -- otherwise stated as having non-trivial priors -- by the inclusion of a term 
$\pi(\nu)$, such that $\mathcal{L}\rightarrow\mathcal{L}\cdot\pi(\nu)$. This will result in a constant term added to per-event significance indicator which looks 
like $\frac{1}{n}\left[\pi(\nu_{\mathcal{B}}) - \pi(\nu_{\mathcal{S}})\right]$, where $\nu_{\mathcal{S}}$ and $\nu_{\mathcal{B}}$ are the values of the nuisance parameters corresponding to the `signal plus background' and `background only' hypotheses, respectively. Although this is important to define the likelihood itself for calculating the significance, this will not change the distribution of $\xi$, so can safely be ignored for the purposes of visualizing the significant excess in the events\footnote{Each nuisance parameter can instead be thought of as an observation (event), in which case, they too have a well defined $\xi$ value, obtained by letting $x_{i} = \nu$ for $i>n$ and $\rho(\nu|H)=\pi(\nu)$. Though not discussed in this note, this could be a useful extension to the method.}. 

\subsection{Limiting case}

Suppose a simple analysis which uses a `histogram' based model for the likelihood. Such an analysis can be thought of as introducing the probability density function, 

\begin{equation*}
    \rho(x_{i}|H) = \begin{cases}
     f_{1} & y_0 < x_{i} \leq y_1 \\
     f_{2}& y_1 < x_{i} \leq y_2 \\
     \text{...} & \\
     f_{m} & y_{m-1} < x_{i} \leq y_m,
    \end{cases}
\end{equation*}
where the $m+1$ boundaries define the `bins' of the histogram and we impose $f_{j}>0$ for all $j$ and $\sum_{i=1}^{m}f_{m}=1$. Let  
the signal and background contributions be represented as $s_{j}$ and $b_j$ such that $f_{j}|_{H_{\mathcal{S}}}=\frac{s_{j}+b_{j}}{S+B}$ and 
$f_{j}|_{H_{\mathcal{B}}}=\frac{b_{j}}{B}$, where $S=\sum_{j=1}^{m}s_{j}$ and $B=\sum_{j=1}^{m}b_{j}$ are the total signal and background over all of the 
bins. Note also then, in this case, $\lambda_{\mathcal{S}}=S+B$ and $\lambda_{\mathcal{B}}=B$. 

In the above case, we find that the per-event significance indicator $\xi$ for an event, which is contained in the $j$-th bin, is given by 

\begin{equation}
    \xi = \ln\left(1+\frac{s_j}{b_j}\right)-\frac{S}{n}  \approx \frac{s_j}{b_j}
\end{equation}
where the approximation can be made in the case that $s_{j}<<b_{j}$ and the total amount of signal $S$ is much smaller than the total number of events $n$. 
Here we can see then, that for this case, we have recovered the typical choice of distribution based on the ratio of the signal to background contribution. Alternatively, we can think of a histogram with $m$ bins as being equivalent to $m$ independent `counting experiments' such that each `event' corresponds to a number of observed events in that bin, $x\rightarrow n_{j}$ and therefore, for each bin $j$, $\rho(n_{j}|H)=1$ and $\mathcal{L}= \frac{\lambda_{j}^{n_{j}}e^{-\lambda_{j}}}{n_{j}!}$. Letting $\lambda_{j}=s_{j}+b_{j}$ and substituting into equation~\ref{eqn:per-event-si}, yields as similar expression $\xi=\ln(1+\frac{s_{j}}{b_{j}})-\frac{s_{j}}{m}$, again recovering the signal to background ratio for $s_{j}<<b_{j}$. It should be noted that this formulation was only possible since the term $b_{i}$ under $H_{\mathcal{B}}$ takes the same value as under $H_{\mathcal{S}}$. In general, this is not true when nuisance parameters are introduced. In this case we find $b\rightarrow b(\nu)$ and in general $\nu$ can take a different value under different hypotheses. Therefore, in the presence of nuisance parameters, the general expression for $\xi$ will \emph{not} be equivalent to $\frac{s}{b}$. Moreover, expression for $\xi$, being more general, also works for unbinned likelihood models as we show in the next sub-section.

\subsection{An example unbinned data search}

Imagine an experimental setup which is able to measure the decay time for some isotope that decays, by some well known process, with a lifetime of $\frac{1}{8}$ ns. Imagine now, that there is some hypothetical a new process (perhaps mediated through some new heavy particle) whose decay time is unknown, other than it should be longer than the standard process - i.e its lifetime is $\frac{1}{\alpha}>\frac{1}{8}$ ns. We can consider searching for this particle by measuring the decay time $x$ of the isotopes, where we limit ourselves to the decay time interval $1\leq x \leq 2$. In some fraction of the events we record, the decay may have proceeded via this new process. This probability density function for this process is well modelled by the sum of two exponential functions, 
\begin{equation}
    \rho(x_i|H_{c}) = \frac{1}{N(c,\alpha)} \left(ce^{-\alpha x_{i}} + (1-c)e^{-8 x_{i}} \right)
\end{equation}
where $0<\alpha<8$. Here, $H_{c}$ represents a family of hypotheses defined by the continuous parameter $0 \leq c\leq1$. The value of 
$N$ serves to ensure that $\int_{1}^{2}\rho(x)dx=1$, for any value of the parameters and is given by $N(c,\alpha)=\frac{c}{\alpha}\left(e^{-\alpha}+e^{-2\alpha}\right)+\frac{1-c}{8}\left(e^{-8}+e^{-16}\right)$.  

\begin{figure}
\begin{center}
\subfigure[][]{\includegraphics[width=0.49\textwidth]{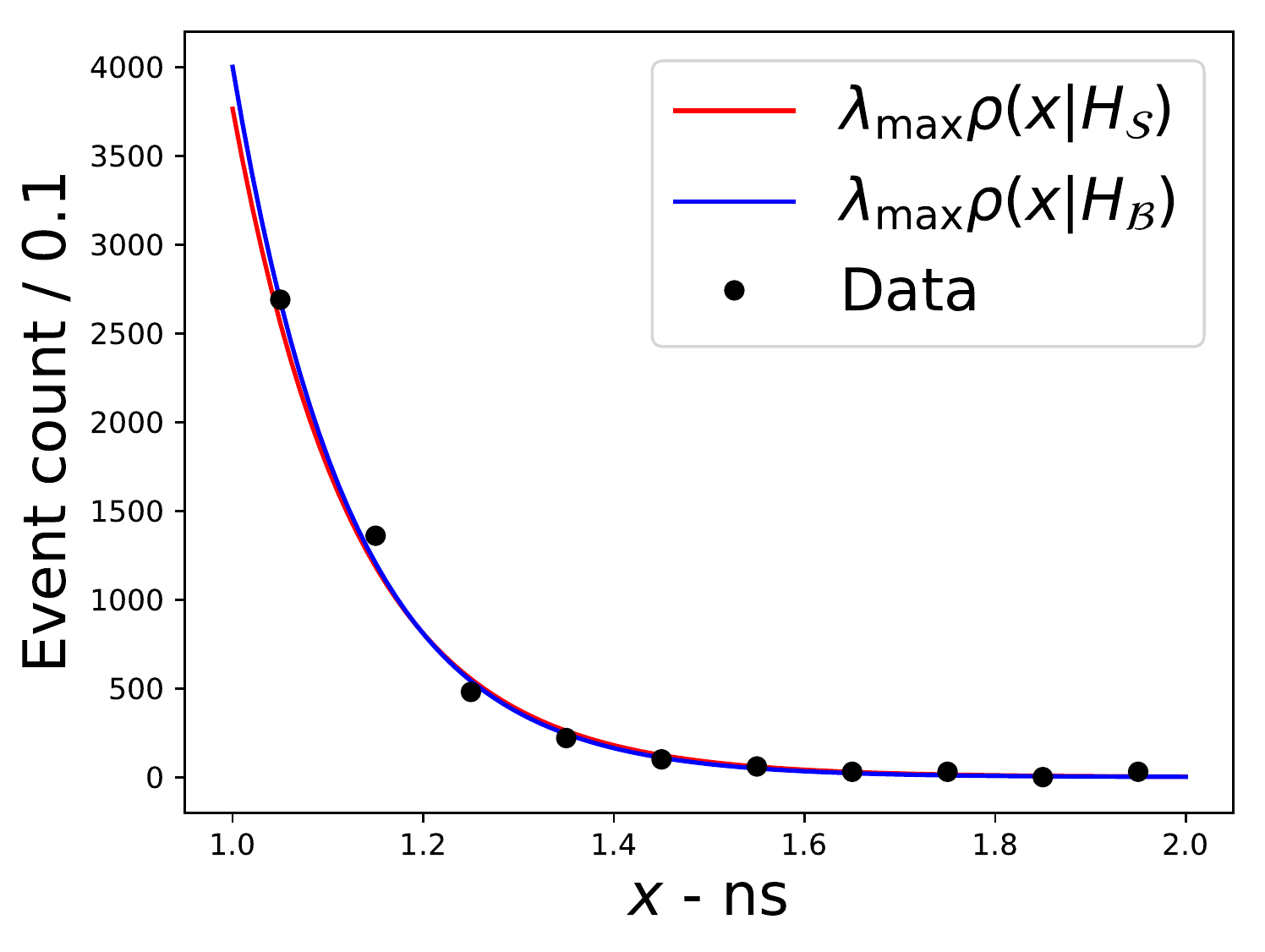}}
\subfigure[][]{\includegraphics[width=0.49\textwidth]{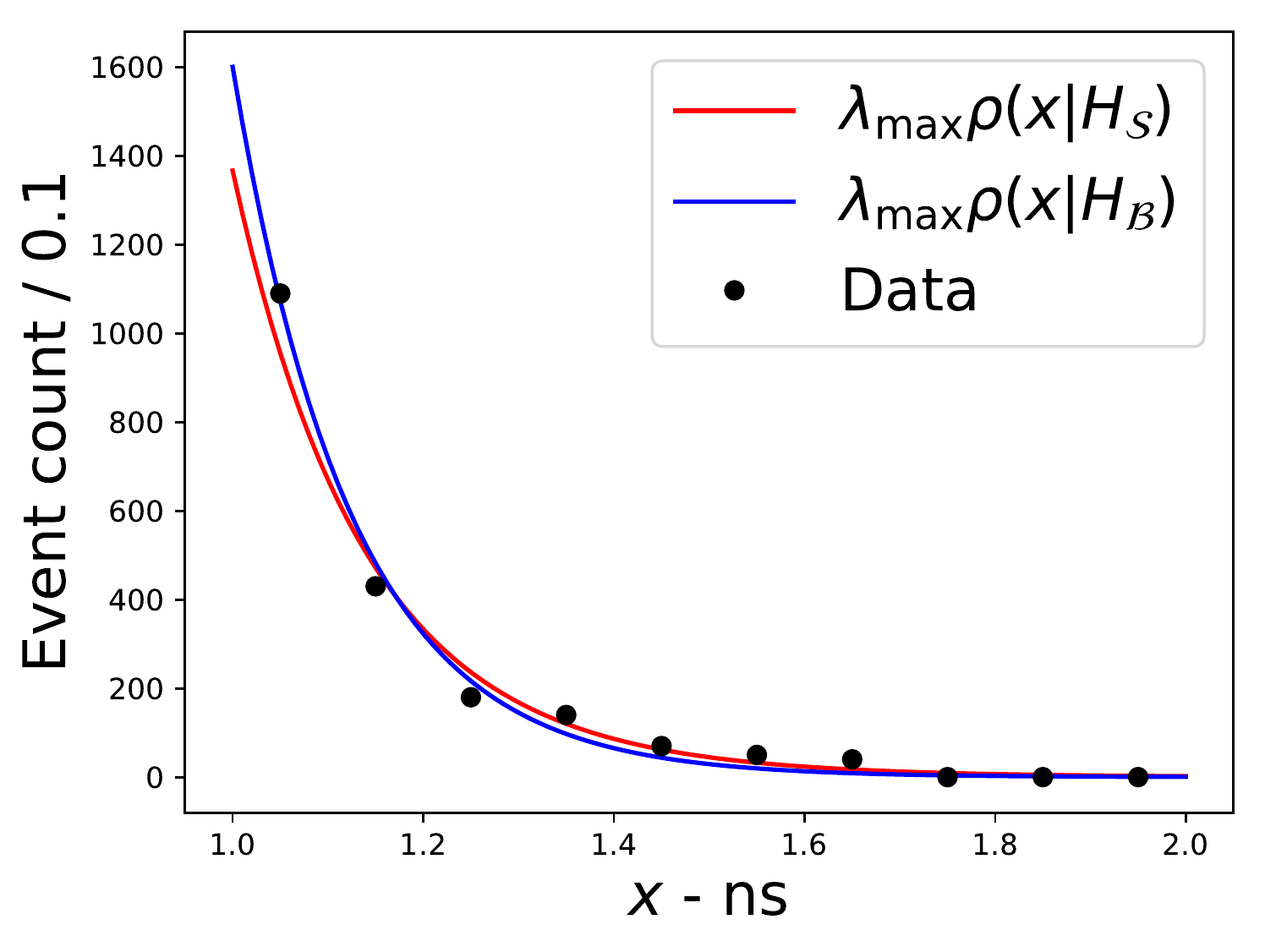}}\\
\subfigure[][]{\includegraphics[width=0.49\textwidth]{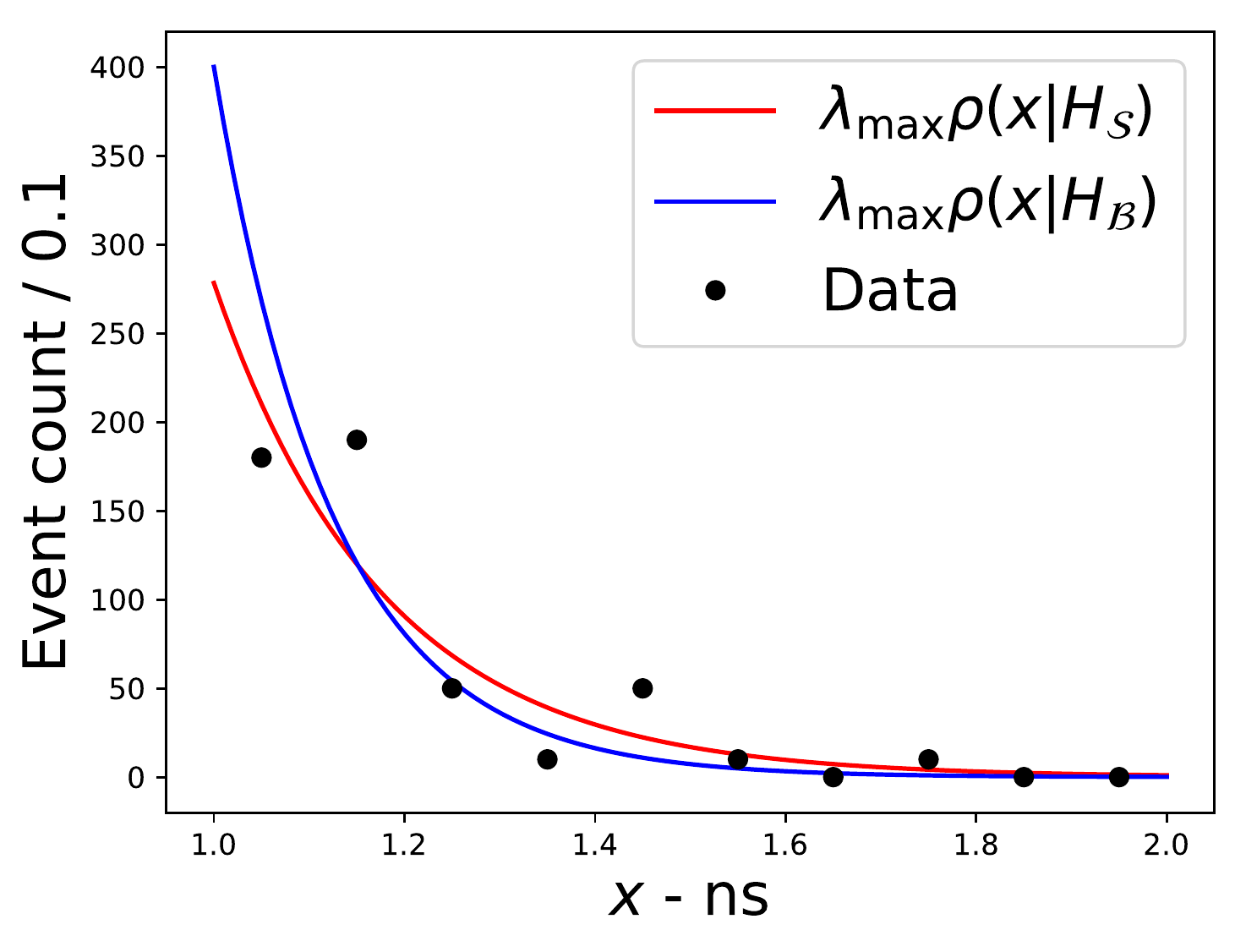}}
\end{center}
\caption{Histograms for the data (black points) and probability density function under the two hypotheses $H_{\mathcal{S}}$ (red line) and  $H_{\mathcal{B}}$ (blue line) for event categories 1 (a), 2 (b) and 3(c). The probability density functions are multiplied by the quantity $\lambda_{\mathcal{S}}$ or $\lambda_{\mathcal{B}}$ to match the normalisation of the data. The binning, 10 bins of width 0.1, used here is only for visualisation purposes.\label{fig:dists_x}}
\end{figure}

It is common to divide the data into different classes, which usually depend on other properties of the events, for example related to the angular distributions of the decay products, in order to improve the sensitivity of the search -- these classes are often referred to as `event categories'. 
As a result, the likelihood is defined as the product of the individual likelihoods from each category $k$, $\mathcal{L}\rightarrow \prod_{k}\mathcal{L}^{k}$. Since events can only enter exactly one category, the per-event significance indicator definition is unchanged. One must however be careful to keep track of 
which event category an event falls in as in general $\rho$, $\lambda$ and $n$ will be different in each category. For this example, we will have three such event 
categories and allow that $\lambda$ and $c$ be different for each of them and hence labelled with the superscript $k$. The unbinned likelihood is defined as;
\begin{equation}
    \mathcal{L}(\alpha.\vec{c},\vec{\lambda}) =  
    \prod_{k=1}^{3} \frac{(\lambda^{k})^{n^{k}}e^{-\lambda^{k}}}{n^{k}!} \prod_{i=1}^{n^k}\frac{1}{N(c^{k},\alpha)} \left(c^{k}e^{-\alpha x_{i}} + (1-c^{k})e^{-8 x_{i}} \right)
\label{eqn:exp_likelihood}
\end{equation}
where the 7 parameters which specify the likelihood has been explicitly shown and $\vec{c}=(c^1,c^2,c^3)$ and $\vec{\lambda}=(\lambda^1,\lambda^2,\lambda^3)$.   
We define $H_{\mathcal{S}}$ at the values of $\alpha,~\vec{c}$ for which $\mathcal{L}$ obtains its maximum value, and denote these values $\alpha_{\text{max}},~\vec{c}_{\text{max}}=(c^{1}_{\text{max}},c^{2}_{\text{max}},c^{3}_{\text{max}})$ -- i.e $L_{\mathcal{S}}=\mathcal{L}(\alpha=\alpha_{\text{max}},\vec{c}=\vec{c}_{\text{max}})$. It is then simple to identify the background only hypothesis $H_{\mathcal{B}}$ with $\vec{c}=0$ -- i.e $\mathcal{L}_{\mathcal{B}}=\mathcal{L}(\vec{c}=0)$.  Likewise, the parameters $\vec{\lambda}$ under $H_\mathcal{S}$ and $H_\mathcal{B}$ are defined as the values which maximize $\mathcal{L}_{\mathcal{S}}$ and $\mathcal{L}_{\mathcal{B}}$, respectively, and are denoted $~\vec{\lambda}_{\mathcal{S}}$ and $~\vec{\lambda}_{\mathcal{B}}$. In this way, the parameters $\vec{\lambda}$ are `nuisance parameters', albeit with trivial constraints, and the test-statistic $t=-2\ln\left(\frac{\mathcal{L}_{\mathcal{B}}}{\mathcal{L}_{\mathcal{S}}}\right)$ is a \emph{profile likelihood ratio}.  From the form of the likelihood in equation~\ref{eqn:exp_likelihood}, it is clear that $\vec{\lambda}_{\mathcal{S}}=\vec{\lambda}_{\mathcal{B}}$ so from hereafter, we will refer to this as $\vec{\lambda}_\text{max}$.
Figure~\ref{fig:dists_x} shows three distributions of events in each category. The fraction of signal $c$ is different for each category, but the value of $c$ in each is unknown. The distributions have been binned into bins of size 0.1 only for the purposes of displaying the events. Additionally figure~\ref{fig:dists_x} shows the probability density functions $\rho(x|H_{c})$, where the free parameters have been set to those obtained by maximizing the likelihood in equation~\ref{eqn:exp_likelihood} under the two hypotheses $H_{\mathcal{S}}$ and $H_{\mathcal{B}}$. Again, 
for the purposes of visualization, these have been multiplied by $\lambda^{k}_{\text{max}}$.  

\begin{figure}
\begin{center}
\includegraphics[width=0.6\textwidth]{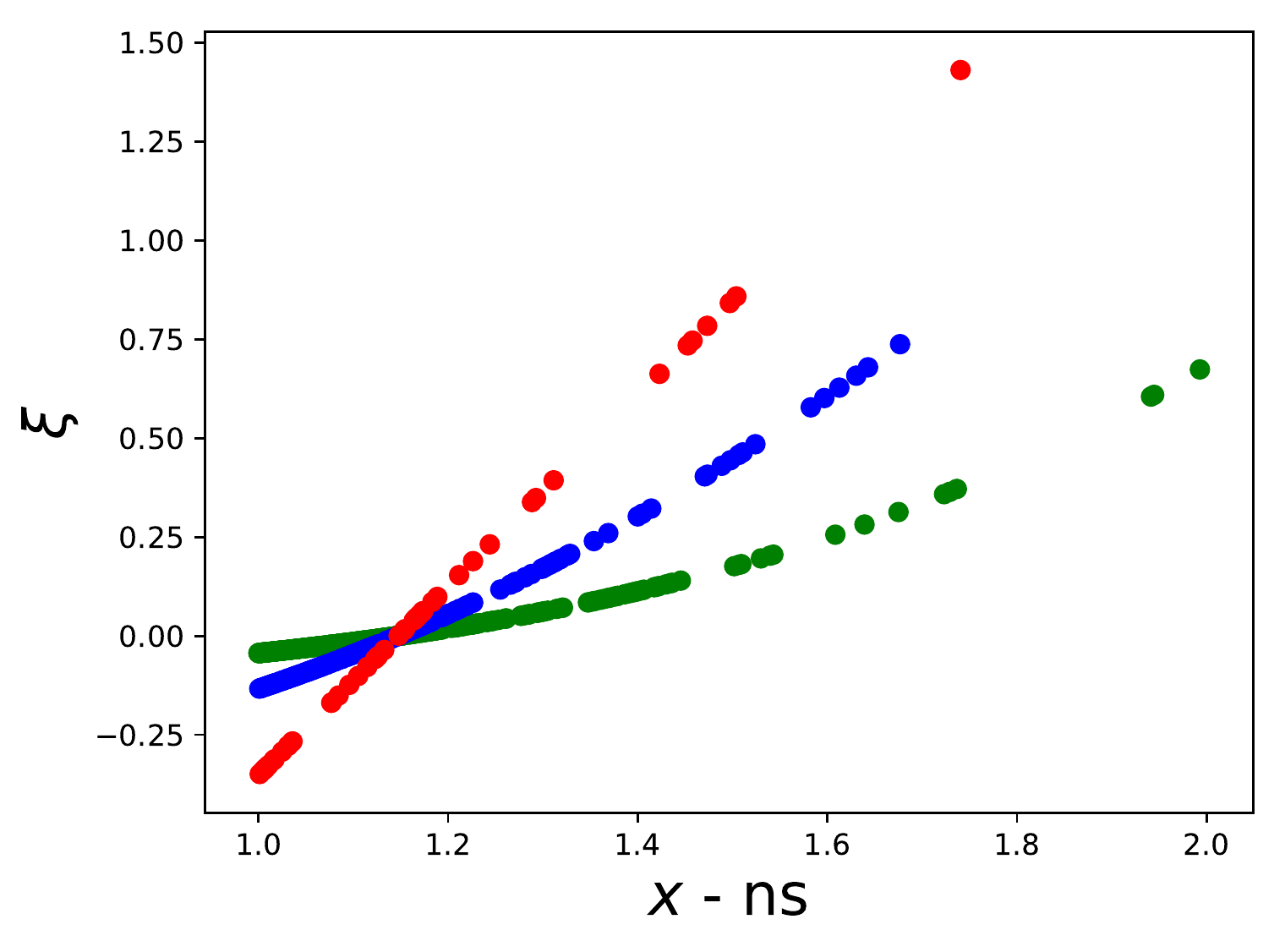}
\end{center}
\caption{Scatter plot of $x$ vs $\xi$ for the events in the three event categories. The events in categories 1, 2 and 3 are shown by the green, blue and red points. \label{fig:xivsx}}
\end{figure}

Substituting into equation~\ref{eqn:per-event-si}, the per-event significance indicator for an event landing in category $k$ is given by;

\begin{equation}
    \xi = \ln \left(c_{\text{max}}^{k}e^{-\alpha_{\text{max}}x}+(1-c_{\text{max}}) e^{-8x}\right) +\beta x  
    - \ln\left(\frac{N(c^{k}=c^{k}_{\text{max}},\alpha=\alpha_{\text{max}})}{N(c^{k}=0)}\right)
\label{eqn:xi_exp}
\end{equation}
The last term in this equation is constant for each value of $x$ in a given category, although in general it will have a different value for each category. Figure~\ref{fig:xivsx} shows the value of $\xi$ and $x$ for each event in each of the three categories. There is a strong correlation between $x$ and $\xi$ and the three categories are clearly differentiated by the  difference in this correlation. Events with a long decay time have a larger contribution to the overall significance of the signal owing to this correlation with $\xi$. It is clear that the greatest correlation is seen in category 3, which also has the largest value of $c_{\text{max}}$. This indicates that, as expected, the category with more signal will generally provide the greatest significance. 

\begin{figure}
\begin{center}
\includegraphics[width=0.6\textwidth]{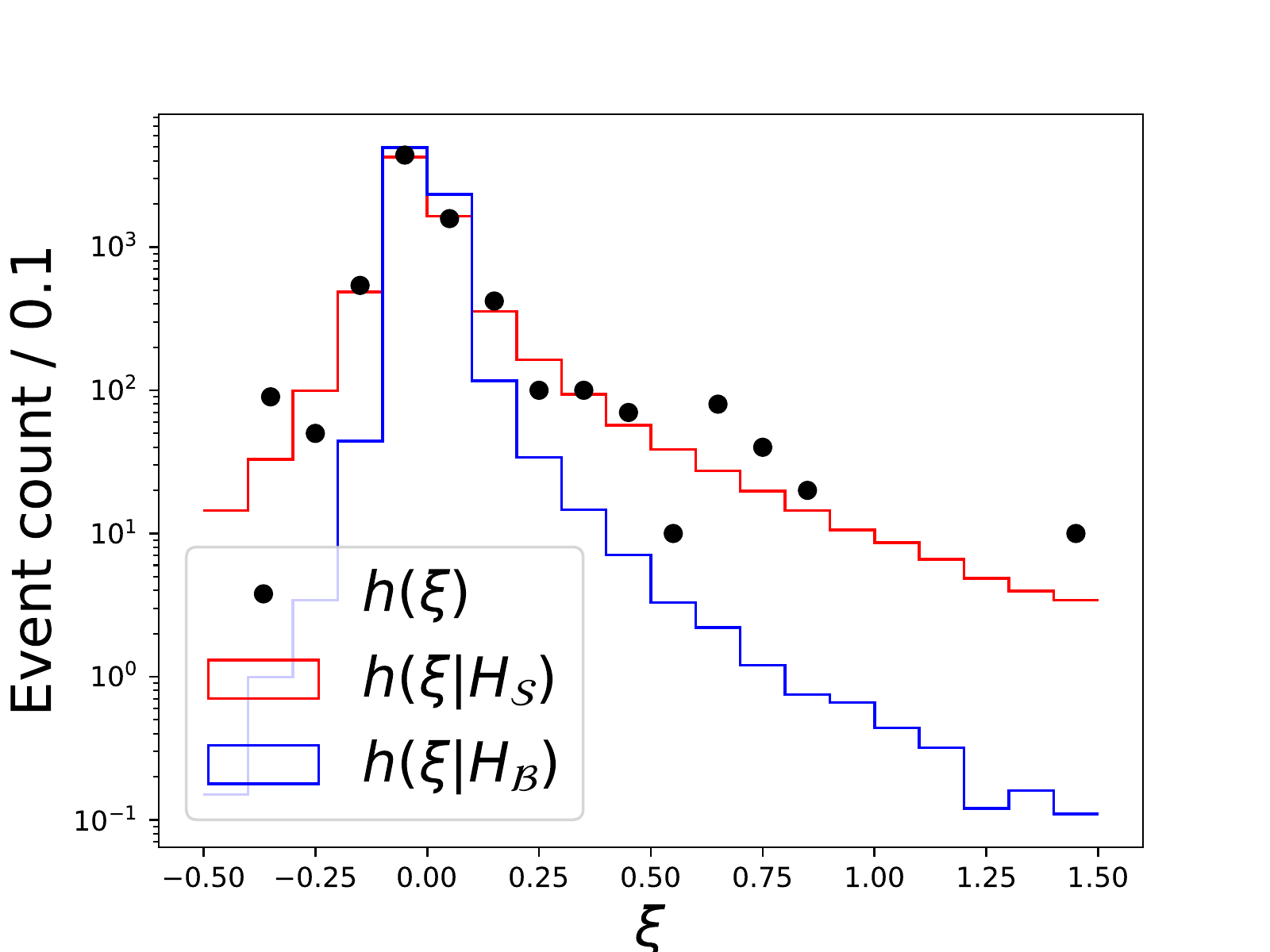}
\end{center}
\caption{Histogrammed distribution of $\xi$ in the data (black points) from all three categories. The distributions expected under the signal plus background hypothesis ($H_{\mathcal{S}}$) and background only hypothesis ($H_{\mathcal{B}}$), generated from pseudo-datasets, are shown in red and blue respectively. \label{fig:xidist}}
\end{figure}

The distribution of $\xi$ is given, as a histogram with bins of size 0.1, in figure~\ref{fig:xidist}. The majority of the events are clustered close to values at 0. These events do not contribute much to the overall significance. Events with large values of $\xi$ however, contribute most to the overall significance.  The expected distribution of events under $H_{\mathcal{S}}$ and $H_{\mathcal{B}}$ are also shown. These distributions are determined by generating pseudo-datasets under each hypothesis from the probability density function $\rho$ in each category, and calculating the distribution of $\xi$ for each dataset. A total of 10,000 pseudo-datasets are generated under each hypothesis so that the weight of each dataset used in the histograms is $10^{-4}$. 

This distribution has the same features as the typical signal to background distributions often shown in that the events with a large contribution to the significance can be identified -- as those with large $\xi$ -- and a comparison of the distribution to the two hypotheses can be made. Distributions such as figure~\ref{fig:xidist} can published to visualise excesses in the data, however, in principle, one could also use the value of $\xi$ to determine which kind of events to include or reject in a particular analysis or design a categorisation scheme for a particular analysis. Of course, anyone using $\xi$ for the purpose of designing an analysis should take care to use pseudo-data (or simulation data) rather than the \emph{real} data to avoid introducing any bias in the results. 

\section{Summary}

In this note, a method of visualising significant events in a hypothesis test has been presented. The per-event significance indicator $\xi$ is derived from a typical form of a 
likelihood, used in particle physics, and can be calculated on an event by event basis. The distribution of $\xi$ indicates the events that contribute significantly to a hypothesis test, such as the observation of an excess of events consistent with a new signal process, from a particular set of events. The quantity $\xi$ is calculated using the probability density under each hypothesis. This means that one is not restricted to the scenario of a `background only' and a `signal plus background' hypothesis test, and naturally extends to situations in which nuisance parameters are included that alter the probability density under each hypothesis. The distribution of $\xi$ can be determined for the case in which the likelihood uses unbinned data, which is not uniquely defined for more typical quantities used to visualise significant events, such as those based on the signal to background ratio. In a simple case of a histogram used as the probability density, with no nuisance parameters affecting the background, the quantity $\xi$ approximates to the ratio of signal to background in a given bin when the signal is small compared to the background. The method has been demonstrated to work for a simple three category unbinned likelihood model. The method can be seen as an alternative way to show distributions of events from many different categories and with different observable values, and indicate which of the events most contribute to a significant result, similar to traditional signal to background distributions.  

\section*{Acknowledgements}
The author would like to thank Louis Lyons for interesting discussions and useful feedback on this note. The author would also like to acknowledge the UK Science and Technologies Facility Council (STFC) who funds his research under the Fellowship grant {\#ST/N003985/1}.

\bibliography{main}
\bibliographystyle{ieeetr}
\end{document}